\documentstyle[aps,preprint,eqsecnum]{revtex}
\def\endignore{}
\def\ignore #1\endignore{}% use to "comment out" text

\ifx\epsffile\undefined
\def\insertfig#1{}% null macro
\else
\def\insertfig#1{{\baselineskip=4pt
\centerline{\epsfxsize=\hsize\epsffile{#1}}}}\fi

\def\eq{\begin{equation}}
\def\eeq{\end{equation}}

\def\eqa{\begin{eqnarray}}
\def\eeqa{\end{eqnarray}}

\def\en#1{(\ref{#1})}

\def\jref#1#2#3#4{#1 {\bf #2}, #3 (#4)}

\def\NPB#1#2#3{\jref{Nucl.\ Phys.}{B#1}{#2}{#3}}

\def\PLB#1#2#3{\jref{Phys.\ Lett.}{#1B}{#2}{#3}}
\def\PR#1#2#3{\jref{Phys.\ Rep.}{#1}{#2}{#3}}
\def\PRD#1#2#3{\jref{Phys.\ Rev.}{D#1}{#2}{#3}}

\def\ie{{\em i.e\/}.}
\def\eg{{\em e.g\/}.}

\def\to{\mathop{\rightarrow}}

\def\myint{\int\mkern-5mu}
\def\sfrac#1#2{{\textstyle\frac{#1}{#2}}}  % small fraction

\def\Dsl{\hbox{\kern.1em/\kern-.7000em$D$}} % D slash

\def\twi{\widetilde}

\def\scr#1{{\cal #1}}

\def\mybar#1{\kern 0.8pt\overline{\kern -0.8pt#1\kern -0.8pt}\kern 0.8pt}
\def\sla#1{\raise.15ex\hbox{$/$}\kern-.57em #1}% Feynman slash
\def\Sla#1{\kern.15em\raise.15ex\hbox{$/$}\kern-.72em #1}% big Feynman slash

\def\roughly#1{\mathrel{\raise.3ex\hbox{$#1$\kern-.75em%
    \lower1ex\hbox{$\sim$}}}}

\def\tr{\mathop{\rm tr}}

\def\det{\mathop{\rm det}}

\def\al{\alpha}

\def\ep{\epsilon}

\def\la{\lambda}
\def\La{\Lambda}
\def\si{\sigma}

\def\th{\theta}

\def\ChPT{\raise.45ex\hbox{$\chi$}PT}

\def\Gc{G^{\rm c}}
\def\sq{/\!\!/}

\begin{document}
\tighten
\preprint{\vbox{
\hbox{MIT-CTP-2440}
\hbox{hep-th/9506098}}}
\title{Varieties of vacua in classical\\
supersymmetric gauge theories}

\author{Markus A. Luty%
\footnote{Address starting January 1996:
Department of Physics, University of Maryland, College Park MD 20742.
E-mail: {\tt luty@ctp.mit.edu}} \\
\medskip
Washington Taylor IV%
\footnote{Address after January 1996:
Department of Physics, Joseph Henry Laboratories, Princeton
University, Princeton NJ 08544.
E-mail: {\tt wati@mit.edu}}
\medskip}

\address{Center for Theoretical Physics \\
Massachusetts Institute of Technology\\
Cambridge, MA 02139\medskip}

\date{June 1995}

\maketitle

\begin{abstract}
We give a simple description of the classical moduli space of vacua
for supersymmetric gauge theories with or without a superpotential.
The key ingredient in our analysis is the observation that the
lagrangian is invariant under the action of the complexified gauge
group $\Gc$.  From this point of view the usual $D$-flatness
conditions are an artifact of Wess--Zumino gauge.  By using a gauge
that preserves $\Gc$ invariance we show that every constant matter
field configuration that extremizes the superpotential is $\Gc$
gauge-equivalent (in a sense that we make precise) to a unique
classical vacuum.  This result is used to prove that in the absence of
a superpotential the classical moduli space is the algebraic variety
described by the set of all holomorphic gauge-invariant polynomials.
When a superpotential is present, we show that the classical moduli
space is a variety defined by imposing additional relations on the
holomorphic polynomials.  Many of these points are already contained
in the existing literature.  The main contribution of the present work
is that we give a careful and self-contained treatment of limit points
and singularities.
\end{abstract}
\pacs{?}

%---------------------------------------------------------------------
\section{INTRODUCTION}
Recently, significant progress has been made in understanding the
structure of 4-dimensional supersymmetric gauge theories.  Building on
earlier work \cite{ads,cern} and using arguments based on symmetry,
holomorphy, and weak-coupling limits, it has been possible to reach
remarkable conclusions about the non-perturbative structure of these
theories \cite{svac,dual}.  Particularly striking results have been
achieved in $N = 2$ theories using these methods \cite{sw}.  One of
the goals of this recent work has been to understand the structure of
the moduli spaces of vacua in supersymmetric gauge theories.  In
ref.~\cite{ads} a methodology was developed for describing the
classical space of vacua in terms of coordinates constructed from
holomorphic gauge-invariant polynomials in the matter fields.
However, in most of the literature this methodology is applied on a
case-by-case basis, with little insight given as to its general
applicability.  The purpose of this paper is to give a simple but
rigorous proof that the moduli space of vacua can be precisely described
in this simple way.
Many of the results we obtain are contained in the existing literature
\cite{wb,witten,mumford}.
The main contribution of the present work is that we give a unified
description of these results which properly takes into account
``fine points'' such as sets of measure zero and
singularities.  These points are important because they often
correspond to physical features such as enhanced gauge symmetry.

Our point of departure is the observation that a supersymmetric gauge
theory with gauge group $G$ is invariant under the complexified gauge
group $\Gc$.  From this point of view, the usual $D$-flatness
conditions can be viewed as a $\Gc$ gauge artifact.  By using a gauge
in which $\Gc$ invariance is preserved, we show that in the absence of
a superpotential {\em every} constant value of the matter fields is
$\Gc$ gauge-equivalent (in an extended sense that we make precise) to
a solution of the $D$-flatness conditions.  This gives the result that
the space of classical vacua is
\eq
\label{vspace}
\scr M_0 = \scr F \sq \Gc,
\eeq
where $\scr F$ is the space of all constant matter field
configurations and the quotient denoted by $\sq$ identifies any $\Gc$
orbits that have common limit points.  This gives a manifestly
holomorphic description of the space $\scr M_0$.  In fact, we can use
this result to prove (using elementary results of algebraic geometry)
that the space of vacua can be described by the set of all
gauge-invariant holomorphic polynomials.  These polynomials form an
algebra generated by a finite number of monomials subject to (finitely
many) defining constraints, as in ref.~\cite{ads}.  That is, $\scr
M_0$ is an algebraic variety.

These results generalize simply to the case where a superpotential is
present.  In that case every constant field configuration that
extremizes the superpotential is $\Gc$ gauge-equivalent (in the
extended sense) to a classical vacuum and the space of classical vacua
is given by eq.~\en{vspace}, where $\scr F$ is the space of stationary
points of the superpotential.  This space $\scr F$ is by definition an
algebraic variety, which is sufficient to show that $\scr M_0$ is a
variety in this case as well.

This paper is organized as follows.  In Section II, we derive our
principal results on the structure of the space of vacua; In Section
III, we give several illustrative examples.  Section IV contains a
discussion of related work and our conclusions.  In the Appendix, we
give a simple proof that the space $\scr M_0$ is a variety.

%---------------------------------------------------------------------
\section{CLASSICAL VACUA}
\subsection{Quotient space}
The lagrangian of a supersymmetric gauge theory can be written%
\footnote{We use the conventions of Wess and Bagger \cite{wb}.}
\eq
\scr L = \myint {\rm d}^2\th\,  {\rm d}^2\mybar\th\, \Phi^\dagger {\rm
e}^V \Phi
+ \left(  \frac 1{4g^2} \myint {\rm d}^2\th\, \tr(W^\al W_\al)
+ \myint {\rm d}^2\th\, W (\Phi) + {\rm h.c.} \right),
\eeq
where $\Phi$ are chiral matter fields transforming in some (in general
reducible) representation of the gauge group $G$, $V$ is a vector
superfield taking values in the Lie algebra of $G$, and $W (\Phi)$ is
a superpotential.  This lagrangian is invariant under a
large group of gauge transformations
\eq
\Phi \mapsto g \cdot \Phi, \qquad
{\rm e}^V \mapsto g^{-1\dagger} {\rm e}^V g^{-1},
\eeq
where $g = {\rm e}^{i\La}$ and $\La$ is a chiral superfield in the Lie
algebra of $G$.  In particular, $\La$ can be a complex scalar, so that
this includes $\Gc$ transformations.  Conventionally, one fixes
Wess--Zumino gauge, which breaks $\Gc$ invariance leaving only
``ordinary'' $G$ gauge invariance.  We will instead use a gauge in
which $V$ takes the form
\eq
V_A = C_A - \th \si^\mu \mybar\th v_{\mu A} + i\th\th \mybar\th\mybar\la_A
- i \mybar\th\mybar\th \th\la_A + \sfrac 12 \th\th \mybar\th\mybar\th D_A,
\eeq
where $A$ is a $G$ adjoint index.  This leaves a residual $\Gc$ gauge
freedom.  It is straightforward to derive the $D$-flatness conditions
in this gauge, which read
\eq
\label{Dflat}
\frac{\partial}{\partial C_A} \left( \phi^\dagger {\rm e}^C \phi \right) = 0,
\eeq
where $\phi$ is the scalar component of $\Phi$.

This immediately shows that any $\phi$ that satisfies the $D$-flatness
conditions~\en{Dflat} for some $C$ is $\Gc$ gauge-equivalent to
the field $\hat\phi = {\rm e}^{C / 2} \phi$, which satisfies
\eq
\label{newDflat}
0 = \frac{\partial}{\partial \hat C_A}
\left. \left(\hat\phi^\dagger {\rm e}^{\hat C} \hat\phi \right)
 \right|_{\hat C = 0}
= \frac{\partial}{\partial \hat C_A} \left. \nu({\rm e}^{\hat C / 2} \hat\phi)
\right|_{\hat C = 0},
\eeq
where
\eq
\nu(\phi) \equiv \phi^\dagger \phi.
\eeq
Eq.~\en{newDflat} is just the usual $D$-flatness condition in
Wess--Zumino gauge.  Since $\nu(\phi)$ is $G$-invariant, we see that
the fields that satisfy these $D$-flatness conditions are precisely
those for which $\nu(\phi)$ is stationary with respect to
$\Gc$.%
\footnote{Essentially the same result is derived in ref.~\cite{wb}.
Similar arguments have been discussed recently by H. Georgi, and by J.
March--Russell (unpublished).} The set of points for which this
condition is satisfied lie on closed $G$ orbits (since $G$ is compact)
that we will refer to as $D$-{\em orbits\/}.

We consider now the case where the superpotential vanishes, and show
that {\em every} constant field configuration $\phi_0$ is $\Gc$
gauge-equivalent to a solution $\hat\phi$ of the
Wess--Zumino gauge $D$-flatness condition eq.~\en{newDflat}.
To make our results
precise, we need a slightly generalized notion of $\Gc$
gauge-equivalence.  We say that two constant field configurations
$\phi_1$ and $\phi_2$ are $\Gc$ equivalent in the {\em extended} sense
if there is a sequence $\{ g_n \}$ of elements in $\Gc$
such that
\eq
\label{limit}
\lim_{n \to \infty} g_n \cdot \phi_1 = \phi_2.
\eeq
In order for this to define an equivalence we must also impose the
same condition with the roles of $\phi_1$ and $\phi_2$ reversed; we
must also impose transitivity, \ie\ $\phi_1$ and $\phi_2$ are
equivalent if there is a $\phi_3$ that is equivalent to both $\phi_1$
and $\phi_2$.  We call the set of all fields that are equivalent in
this sense to a field $\phi$ the {\em extended} $\Gc$ {\em
orbit} of $\phi$.  These definitions are physically sensible because
any gauge-invariant function takes the same value on all
the field configurations in an extended orbit,  so that the points of
such an orbit are physically indistinguishable.

With these definitions, the result to be proven can be concisely
stated: every extended $\Gc$ orbit contains a $D$-orbit.  This
immediately implies that the space of classical vacua is given by
\eq
\label{modresult}
\scr M_0 = \scr F \sq \Gc,
\eeq
where $\scr F$ is the space of all constant matter field
configurations, and the extended quotient by $\Gc$ is defined using
the equivalence defined above.  This result is intuitively satisfying
since it is closely analogous to the result for non-supersymmetric
theories that  (in a theory with no potential) every
constant field configuration lies in a gauge equivalence class of
vacua.

The proof of this assertion is extremely simple.
Fix an arbitrary $\phi_0$.
Since the function $\nu(\phi)$ is positive
semidefinite and is less than or equal to $\nu(\phi_0)$ only on a
compact ball in $\phi$-space, it must take on a minimum value at some
point in the closure of the ordinary $\Gc$ orbit that contains
$\phi_0$.  Thus, there is a $\hat\phi$ such that
\eq
\hat\phi = \lim_{n \to \infty} g_n \cdot \phi_0,
\eeq
which minimizes $\nu$ on the closure of the orbit.
Clearly, $\hat\phi$ lies in the extended $\Gc$ orbit containing $\phi_0$.
Furthermore, $\nu(\hat\phi)$ must be stationary with respect to $\Gc$
transformations, since otherwise we could construct a different
sequence that converges to a new value of $\hat\phi$ with smaller
$\nu(\hat\phi)$ by making a $\Gc$ transformation of the original
sequence.  Thus, $\hat{\phi}$ is in a $D$-orbit.%
\footnote{A different argument for essentially the same conclusion is
given in ref.~\cite{wb}.}

This result makes it intuitively clear why the space of classical
vacua can be parameterized by the set of gauge-invariant holomorphic
polynomials in the fields $\phi$, as advocated in ref.~\cite{ads}.
Such polynomials are constant on extended $\Gc$ orbits, and it seems
natural that there are ``enough'' polynomials to distinguish any two
distinct extended orbits.  In the appendix, we show that this
intuition can be made rigorous using fairly elementary
results from algebraic geometry.  We prove that the space $\scr M_0$
has as coordinates a set of gauge-invariant polynomials subject to
finitely many defining relations.  In the language of algebraic
geometry, $\scr M_0$ is the algebraic variety defined by the ring of
all invariant polynomials on $\Phi$.
%This variety is uniquely defined by the algebra of gauge-invariant
%polynomials in $\phi$.

The argument above can be extended immediately to the case where there
is a superpotential present.  In that case, the fields must extremize
the superpotential
\eq
\label{fdef}
R_j(\phi) \equiv \frac{\partial W(\phi)}{\partial \phi_j} = 0
\eeq
as well as satisfying the $D$-flatness conditions.  It is easy to see
that if any point in an extended $\Gc$ orbit satisfies (\ref{fdef})
then all other points in that extended orbit also satisfy this
equation.  We can thus simply restrict $\phi$ to satisfy
eq.~\en{fdef} and proceed as above.  The result is that the space of
vacua is given by eq.~\en{modresult}, where $\scr F$ is the space of
fields that extremize the superpotential.
(See also ref.~\cite{wb}.)

It is straightforward to describe the classical moduli space of vacua
in theories with a superpotential as a variety.  The results proven in
the appendix show that the moduli space can be parameterized by the
gauge-invariant polynomials on the set of fields that extremize the
superpotential.  This means that in addition to the defining
relations, there are extra relations on the polynomials stating that
any gauge-invariant combination of the $R$'s defined in eq.~\en{fdef}
with the $\phi$'s must vanish.  We will give an example of this
construction in Section III.

%---------------------------------------------------------------------
\subsection{Observations on orbit structure}
We now collect some observations about the structure of extended
orbits.  The main results of this paper do not depend on these
observations, but we include them to clarify the significance of
the extended $\Gc$ orbits.
We first show that there is exactly one $D$-orbit in
every extended orbit.  This shows that the classical moduli
space can be precisely identified with the set of solutions to the
Wess--Zumino gauge $D$-flat conditions with points in the same $G$
orbit identified, and provides a simple connection between our
approach and the conventional treatment.  We then discuss the
relationship between extended orbits and points of enhanced symmetry.
We show that in any extended orbit that contains more than one
ordinary $\Gc$ orbit, points in the $\Gc$ orbit containing the
$D$-orbit have more gauge symmetry than points in other orbits of the
same extended orbit.

To show that there is a unique $D$-orbit in every extended orbit, we
begin by showing that every stationary point $\hat\phi$ of $\nu(\phi)$ on an
ordinary $\Gc$ orbit $O$ lies in a $D$-orbit which is a global
minimum of $\nu$ in $O$.
Along any exponential curve
\eq
\label{limcurve}
\phi(t) = {\rm e}^{tC / 2} \phi_0,
\eeq
because $\nu$ is positive semidefinite we have\footnote{We thank H.
Georgi for this observation.}
\eq
\label{eq:second}
\frac{\partial^2}{\partial t^2} \nu(\phi(t)) = \nu(C \phi(t)) \ge 0.
\eeq
Eq.~\en{eq:second} can vanish for finite $t$ only if $C \phi(t) = 0$,
which is only possible when $\phi (t)= \phi_0$ for all $t$.
Every element of $\Gc$ can be written in the form
\begin{equation}
g = {\rm e}^{C} \cdot u
\end{equation}
where $C$ is Hermitian and $u \in G$.  Thus, every point in $O$ can be
reached by an exponential curve starting at a point in the same
$D$-orbit as $\phi_0$, and $\nu$ is monotonically increasing along every
such curve.  This proves that the $D$-orbit is a global
minimum of $\nu$ in $O$.  In fact, because the $D$-orbit is compact, it
is not hard to see that the set of points in $O$ where $\nu$ is less
than or equal to any fixed number $x$ is a compact set.  This implies
that any limit of a sequence in $O$ which does not lie in $O$  would
have a divergent value of $\nu$, which implies that $O$ is a closed
orbit containing all its limit points.

We cannot immediately conclude from this that $\hat\phi$ minimizes $\nu$
on the extended orbit $X$, since there are in general directions in $X$
which do not correspond to $\Gc$ transformations.%
\footnote{As an example of the type of difficulty which may arise, we
mention that there are examples where a point $\hat\phi$  is the
limit of a sequence of points $g_n \cdot \phi_0$, and yet there is
no exponential curve ${\rm e}^{tC}\phi_0$ that approaches $\hat\phi$.}
We can however use the fact that the action of $\Gc$ is
algebraic to show that every extended orbit contains a unique
$D$-orbit.  We have shown that every ordinary $\Gc$ orbit which
contains a $D$-orbit is closed.  The proof of statement (i) in the
Appendix shows that for any two disjoint closed sets which are
invariant under $\Gc$, there exists a gauge invariant polynomial
which takes different values on the two sets.  Thus,
two distinct closed orbits cannot lie in a single extended orbit
This clearly implies that each extended orbit contains a unique
$D$-orbit.

Note that the above proof does not hold when the
group $G$ is not compact.
A simple example is an abelian theory with relatively irrational charges.%
\footnote{We thank A. Nelson for suggesting this example.}
In this case, the gauge group $G$ is not compact, and a single
extended orbit contains multiple $D$-orbits.

We now discuss the connection between extended orbits and enhanced
gauge symmetry.  On any ordinary $\Gc$ orbit, the invariant subgroup
of $\Gc$ is the same (up to conjugation) at all points on the orbit.
However, in an extended orbit $X$ the $\Gc$ orbit containing the
$D$-orbit contains points with more gauge symmetry than the points in
other orbits in $X$.  This can be seen intuitively by noting that if a
sequence of points in one ordinary $\Gc$ orbit $O$ approaches a point
in another orbit $\hat O$ then the direction in which the limit is
approached corresponds to an extra invariance of the limit point.  Due
to the complications mentioned above it is easier to make this result
precise using algebraic arguments.  As noted in the Appendix, every
orbit can be written as a finite union and intersection of algebraic
sets.  This implies that in the situation above, since $\hat{O}$ must
be contained in the closure of $O$, the dimension of $\hat O$ must be
strictly smaller than that of $O$.

Based on this one might suppose that every extended orbit
corresponds to a vacuum with enhanced gauge symmetry.  However, in
theories with no flat directions there is a single extended $\Gc$
orbit which contains multiple ordinary $\Gc$ orbits, but there is clearly no
extra gauge symmetry.  One might also conjecture that one can identify
points with extra gauge symmetry from the singularity structure of the
resulting variety, but we will give several examples which show that
this is not possible.

%---------------------------------------------------------------------
\section{EXAMPLES}
\subsection{Supersymmetric QED}
Our first example is supersymmetric QED, a theory with gauge group $G
= U(1)$, a matter field $Q$ with charge $1$, and a matter field
$\twi Q$ with charge
$-1$.  We use this simple example to illustrate the structure of the
extended $\Gc$ orbits.  The classical moduli space in this case is
parameterized by
\eq
A \equiv Q \twi Q
\eeq
so the moduli space can be identified with the set of all complex
numbers $\bf C$.

To understand the $\Gc$ orbit structure, note that $U(1)^c$ is simply
the multiplicative group of non-zero complex numbers.
The action of $\Gc$ in this case is therefore
\eq
(Q, \twi Q) \mapsto (\al Q, \al^{-1} \twi Q)
\eeq
with $\al \ne 0$.  The extended orbit corresponding to a value $A \ne
0$ is the set of points
\eq
(Q, \twi Q) = (q, A / q)
\eeq
with $q \ne 0$, which all lie on an ordinary $\Gc$ orbit.
On the other hand, the extended orbit with $A = 0$ contains three ordinary
$\Gc$ orbits:
\eq
(Q, \twi Q) = (q, 0),\ (0, \twi q),\ \hbox{or}\ (0, 0)
\eeq
with $q, \twi q \ne 0$.
The orbit $(0,0)$ is a limit point of the other two orbits.
Note that the point $A = 0$ is a point of enhanced symmetry,
but the moduli space is completely non-singular there.

The structure of the classical moduli space in this theory is
extremely simple, but it illustrates many of the features we have
described above.  Generic extended $\Gc$ orbits ($A \ne 0$) contain a
single ordinary $\Gc$ orbit which contains a single $D$-orbit.  At
points of enhanced symmetry ($A = 0$), the extended orbit contains
multiple ordinary $\Gc$ orbits, of which only one contains a
$D$-orbit.  In this case, the $\Gc$ orbit which contains the $D$-orbit
has enhanced gauge symmetry, while the other orbits do not.  In this
extended orbit the orbit containing a $D$-orbit is closed and contains
all its limit points, while the remaining orbits contain curves
approaching the $D$-orbit.

%---------------------------------------------------------------------
\subsection{A $U(1) \times U(1)$ model}
We now consider a chiral theory with a more interesting classical
moduli space.
The gauge group is $U(1) \times U(1)$, and the matter fields have charges
\eq
Q \sim (2, 0), \quad
R \sim (-2, 1), \quad
S \sim (1, -1), \quad
T \sim (-1, 0).
\eeq
The gauge-invariant polynomials are generated by
\eq
A = Q R^2 S^2, \quad
B = Q T^2, \quad
C = QRST,
\eeq
which satisfy the defining relation
\eq
AB = C^2.
\eeq
This simple two-dimensional variety is an example of a {\em quadric
surface\/}.  The only singular point on this variety is the point $A =
B = C = 0$.  (This can be seen by noting that when $A \neq 0$ the
variables $(A,C)$ are good coordinates and when $B \neq 0$, $(B,C)$
are good coordinates.)

This classical moduli space has a one-parameter family
of nontrivial extended orbits.
For every $B \ne 0$, there is an
extended orbit with coordinates $A = C = 0$ which contains three
ordinary $\Gc$ orbits
\eq
(Q, R, S, T) = (B / t^2, 0, s, t),\ (B / t^2, r, 0, t),\ \hbox{or}\
(B / t^2, 0, 0, t).
\eeq
where $r, s, t \ne 0$.  The orbit with $R = S = 0$ contains a
$D$-orbit which has enhanced gauge symmetry.
(The second $U(1)$ is unbroken.)
Note that the variety is
not singular on the vacua corresponding to these orbits.  It is also
amusing to note that the extended orbit structure is not symmetric under
interchanging $A$ and $B$, even though the variety is.  This
again illustrates that the presence of enhanced gauge symmetry cannot
in general be detected from the  structure of the variety.

%---------------------------------------------------------------------
\subsection{Supersymmetric QCD with $N_F = N$}
Our final example illustrates how one can obtain a simple description
of the moduli space in the presence of a superpotential.  Consider
supersymmetric QCD, $SU(N)$ gauge theory with chiral superfields
$Q^{aj}$ ($j = 1, \ldots, N_F;\ a = 1, \ldots, N$) in the fundamental
representation and chiral superfields $\twi Q_{ak}$, ($k = 1, \ldots,
N_F$) in the antifundamental representation.  We consider here the
special case $N_F = N > 2$.  According to the discussion above (or
from ref.~\cite{ads}), the classical space of vacua can be
parameterized by the variables
\eqa
M^j{}_k &\equiv& Q^{aj} \twi Q_{ak}, \nonumber\\
B &\equiv&
\frac 1{N!} \ep_{a_1 \cdots a_N} \ep_{j_1 \cdots j_N}
Q^{a_1 j_1} \cdots Q^{a_N j_N}, \\
\twi B &\equiv&
\frac 1{N!} \ep^{a_1 \cdots a_N} \ep^{k_1 \cdots k_N}
\twi Q_{a_1 k_1} \cdots \twi Q_{a_N k_N}, \nonumber
\eeqa
subject to the constraint
\eq
B \twi B = \det(M).
\eeq
We wish to consider the theory in the presence of a superpotential
\eq
W = b B + \twi b \twi B
\eeq
with $b, \twi b \ne 0$.  (We do not add a mass term.)  According to
the discussion in the main part of the paper, the moduli space in the
presence of the superpotential is given by imposing the additional
constraints that all gauge-invariant polynomials which can be
constructed from
\eq
R_{aj} \equiv \frac{\partial W}{\partial Q^{aj}}, \qquad
\twi R^{ak} \equiv \frac{\partial W}{\partial \twi Q_{ak}}
\eeq
vanish.
We must therefore impose
\eqa
\label{cone}
R_{aj} Q^{ak} &=& 0, \\
\label{ctwo}
\twi R^{aj} \twi Q_{ak} &=& 0, \\
\label{cthree}
R_{aj} \twi R^{ak} &=& 0, \\
\label{cfour}
\ep^{a_1 \cdots a_N} R_{a_1 j_1} \cdots R_{a_r j_r}
\twi Q_{a_{r+1} k_{r+1}} \cdots \twi Q_{a_N k_N} &=& 0,
\qquad (r = 1, \cdots, N) \\
\label{cfive}
\ep_{a_1 \cdots a_N} \twi R^{a_1 k_1} \cdots \twi R^{a_r k_r}
Q^{a_{r+1} j_{r+1}} \cdots Q^{a_N j_N} &=& 0, \qquad (r = 1, \cdots, N).
\eeqa
Expressed in terms of the $A$'s and $B$'s, eqs.~\en{cone} and \en{ctwo} give
\eq
B = \twi B = 0.
\eeq
The left-hand-sides of eqs.~\en{cfour} and \en{cfive} for $r > 1$ have
non-zero baryon number and therefore vanish when expressed in terms of
the $M$'s and $B$'s when $B = \twi B = 0$.  For $r = 1$, we obtain
\eqa
\label{ccone}
\ep_{j_1 \cdots j_N} M^{j_2}{}_{k_2} \cdots M^{j_N}{}_{k_N} &=& 0, \\
\label{cctwo}
\ep^{k_1 \cdots k_N} M^{j_2}{}_{k_2} \cdots M^{j_N}{}_{k_N} &=& 0.
\eeqa
Eq.~\en{cthree} gives the constraint
\eq
\ep_{j_1 \cdots j_N} \ep^{k_1 \cdots k_N}
M^{j_2}{}_{k_2} \cdots M^{j_N}{}_{k_N} = 0,
\eeq
which is clearly implied by eqs.~\en{ccone} and \en{cctwo} above.
Thus, the classical moduli space is the space of $M$'s subject to
eqs.~\en{ccone} and \en{cctwo}.
To understand the meaning of these constraints, note that we can use the
$U(N)_+ \times U(N)_-$ global symmetry of the model to diagonalize $M$.
It is then easy to see that eqs.~\en{ccone} and \en{cctwo} impose the same
constraint, namely that the rank of $M$ be at most $N - 2$.
This is therefore the defining constraint of the classical moduli
space.

\section{DISCUSSION}
We have shown that in classical supersymmetric gauge theories, every matter
field $\phi$ that extremizes the superpotential is related by a (limit of a)
complex gauge transformation to a vacuum.
Furthermore, we have proven that the space $\scr M_0$ of classical vacua
has a natural structure as an algebraic variety.

There is a related approach to describing the classical space of vacua
that follows from the observation that the usual gauge-fixed
$D$-flatness equations precisely describe the symplectic reduction of
$\scr F$ by $G$.  This point of view was used by Witten \cite{witten}
to discuss $N = 2$ abelian gauge theories in two dimensions.  The
symplectic quotient of a complex space by $G$ is closely related to
the holomorphic quotient by $\Gc$, which is the natural domain of
geometric invariant theory.  Our result in IIB connecting the space of
extended orbits to the space of $D$-orbits makes this connection
precise for the cases of physical interest.  The approach taken in the
present paper has the virtue that the quotient space structure emerges
naturally and directly as a result of the underlying complexified
gauge symmetry.  Furthermore, the explicit description of the
structure of extended orbits allows us to rigorously describe the
quotient space as an algebraic variety without the application of
sophisticated mathematical theorems.

Several aspects of the picture that we have presented in this paper
have also been considered by others.
A closely related argument for the existence of fields  minimizing
the $D$-term potential appears in ref.~\cite{wb}.
A local holomorphic description
of the space of vacua was given in ref.~\cite{pr}.  During the
completion of the present work, we learned that J. March--Russell has
also studied the relationship between the $D$-flat equations and $\Gc$
orbits, and that H. Georgi has also recently made progress in this direction.

It should be emphasized that the descriptions of $\scr M_0$, both as
an extended quotient space and as an algebraic variety, give the
precise structure of the space of vacua including isolated special
points and singularities.
This is important, since such ``fine points'' often have physical
significance.
For example, we have shown that there is a close connection between
vacua with enhanced gauge symmetry and orbits of the complexified gauge
group which do not contain all their limit points.
At such vacua, the moduli space is often singular.
These singularities continue to play an important role in the quantum theory,
where they may change structure or disappear by being blown up \cite{svac}.
It seems natural to pursue a further understanding
of the classical and quantum moduli spaces of vacua using this
geometrical point of view.

%--------------------------------------------------------------------
\section*{APPENDIX: PROOF THAT $\scr M_0$ IS A VARIETY}

\newcommand{\ccc}{ {\bf C}}
\newcommand{\ring}[2]{{\bf C} [{#1}_1, \ldots,{#1}_{#2}]}
\newtheorem{thm}{Theorem}

In this appendix we give a proof that for any gauge group $G$ and
matter fields $\phi$ in any representation of $G$ the classical moduli
space $\scr F \sq \Gc$ can be parameterized by a finite set of
gauge-invariant polynomials $P_a(\phi)$ subject to a finite number of
relations.  Specifically, we show that $\scr F \sq \Gc$ is the
natural algebraic variety associated with the ring of {\em
all} invariant polynomials in $\phi$.  The proof is valid when there
is a superpotential present, in which case the space $\scr F$ is the set
of values for the fields $\phi$ at which the superpotential is
stationary.  The presence of a superpotential simply imposes additional
relations on the polynomials $P_a$, as described in Section IIA.
In fact, the result holds for any theory where $\scr F$ can be
described as a variety in terms of a set of fields transforming
linearly under $G$ and satisfying a set of algebraic equations.

The proof we give here is essentially a distillation of results
contained in a related proof in ref.~\cite{mumford}.  Our goal in
presenting this proof here is to make this result accessible to the
physics community by giving a self-contained derivation using fairly
elementary methods.  We will use the language of algebraic geometry
but we will only use a few basic definitions and results from this
subject.  We begin by reviewing those concepts and results that we
will use, all of which can be found on the first few pages of any
standard textbook (such as Hartshorne
\cite{hartshorne}).

The set $A$ of points $(x_1,\ldots,x_n)$ in the complex vector space
${\bf C}^n$ satisfying a system of polynomial equations
$f_\alpha(x_1,\ldots,x_n)= 0$ is called an {\em algebraic set\/}.  The
algebraic sets define a special topology on $\ccc^n$ called the {\em
Zariski topology\/}.  In the Zariski topology the closed sets are the
algebraic sets.  Open sets are those sets whose complement is closed.
All of the usual statements of topology hold in the Zariski topology;
{\em e.g.}, the intersection of a finite number of closed sets is
closed, {\em etc\/}.  We will distinguish sets closed in the Zariski
topology from sets closed in the usual topology by using the terms
Z-closed and closed respectively.  It is easy to see that every
Z-closed set is closed and thus that every Z-open set is open.  A {\em
constructable} set is a set which can be constructed from Z-closed and
Z-open sets with a finite number of operations such as unions or
intersections.  Constructable sets have the nice property that every
point in their Z-closure is also in their closure.  (This can be
shown, for example, by first proving the assertion for an algebraic
curve (1-dimensional variety) and then proceeding by induction,
reducing the dimension of the initial variety by one by imposing the
constraint that an additional equation vanishes.)

Associated with every algebraic set $A$ there is a ring of
polynomials $I(A)$ that consists of all polynomials in the variables
$x_i$ that vanish at all points of $A$.  $I(A)$ is an {\em ideal}
(invariant subring) of the ring $\ring{x}{n}$ of all polynomials in
the $x_i$'s.  The {\em Hilbert basis theorem} states that $I(A)$ always has
a finite number of generators, so that $A$ can always be described as
the set of points on which a finite set of polynomials vanishes.

An algebraic set $A$ is {\em irreducible} when it cannot be written as
a union $A = B \cup C$ of two algebraic sets that are proper subsets
of $A$.  An irreducible algebraic set is an {\em affine variety}.  A
Z-open subset (with respect to the induced topology) of an affine
variety is a {\em quasi-affine variety\/}.  We refer to both as simply
varieties.  Every variety $A$ has associated with it a ring $R(A)$ of
rational functions without poles on $A$.  It can be shown that for an
affine variety $A$, $R(A)$ is just $\ring{x}{n} / I(A)$, the
polynomials in the $x_i$'s subject to the relations defined by $I(A)$.
The essential point of algebraic geometry is that all the geometric
information about the variety $A$ is encoded in the algebraic
structure of the ring $R(A)$.  Thus, in algebraic geometry the
fundamental objects are commutative rings rather than geometric
objects.

The simplest example of how $R(A)$ encodes geometric information about
$A$ is given by the algebraic description of points in
$A$.  From the above definitions, it is clear that any Z-closed subset
$B$ in $A$ can be associated with an ideal $I(B) \supset I(A)$.  Thus,
$I(B)$ naturally corresponds to an ideal $I(B) / I(A)$ in $R(A)$.
Conversely, every nontrivial ideal $I$ of $A$ (an ideal that is
neither $\{ 0 \}$ nor $A$) can be associated with a closed, non-empty
algebraic set, the {\em zero set} $Z(I)$ of $I$.  Using another
theorem due to Hilbert (the {\em Nullstellensatz}), it can be shown
that the points in $A$ are in 1-1 correspondence with the ideals $I
\subset R(A)$ that are {\em maximal} in the sense that there exists
no larger ideal $I' \supset I$ other than $I' = R$.

%In the following, we will make use of the fact that every irreducible
%Z-closed set $A$ in ${\bf C}^n$ is connected.  This follows from the
%fact that a function that is constant on each connected component of
%a Z-closed set $A$ is in $R (A)$.  Therefore every connected component
%of $A$ is a Z-closed set and $A$ cannot be irreducible unless it has
%only one connected component.
%
An {\em algebraic map} (or {\em morphism}) is a map from a variety
$A \subset\{(x_1, \ldots,x_n)\}$ to another variety
$B \subset\{(y_1, \ldots,y_m)\}$ that can be described by writing the
$y_i$'s as rational functions of the $x_i$'s with denominators that
are nonvanishing everywhere on $A$.
Such a map gives rise to a ring homomorphism $R(B) \to R(A)$.
It can be shown that the image of a variety under an algebraic map is
always a constructable set.

This concludes our brief review of concepts from algebraic geometry.
In terms of this language, the statement that we wish to prove is the
following:

\medskip\noindent{\bf Theorem\ }
{\em Given a group $\Gc$ acting on a variety $A$, there is a 1-1
correspondence between $A \sq \Gc$ and the set of points in the affine
variety $A^G$ defined by the ring $R_G$ of $G$-invariant elements in
$R =R(A)$.}

\medskip
\noindent
We are making the technical assumptions (which are always valid in the
relevant physical theories) that $A$ is an affine variety in a complex
vector space ${\bf C}^n$ on which $\Gc$ acts linearly, and that $G$ is
the product of a semi-simple Lie group with a torus $U(1)^k$.  Thus,
$\Gc$ is itself a variety (a so-called {\em algebraic group}), and the
action of $\Gc$ on $A$ is described by an algebraic map $\tau : G \times
A \to A$.  The $\Gc$ orbits in $A$ are the image under $\tau$ of $G
\times \{p\}$ where $p$ is a point in $A$; therefore each orbit is a
constructable set.   (In fact, it can be shown that each orbit is a
variety but we will not need that condition.)

Implicit in the statement of the theorem is the result that $A^G$ is
an affine variety.  This follows from the fact that $R_G$ is finitely
generated, which is a consequence of the Hilbert basis theorem and the
fact (used and proven in the proof below) that every ideal $I \subset
R_G$ generates an ideal $M$ in $R$ with $M \cap R_G = I$.

It will be convenient for us to think of $A$ as lying in the complex
vector space with coordinates $x_1, \ldots, x_n$.  We can then take a
set of generators for $R_G$ to be some set $P_1, \ldots, P_\ell$ of
$G$-invariant polynomials in the $x_i$'s.  There is a natural map
$\pi: A \to A^G$ that can be defined by simply evaluating the
polynomials $P_a$ at a point $x \in A$.  Since the polynomials are
invariants, this map is constant on orbits of $\Gc$, so for any point
$p \in A^G$ the preimage $\pi^{-1} (p)$ is a union of disjoint orbits.
Furthermore, by continuity $\pi$ must be constant on extended $\Gc$
orbits in $A$, so it induces a well-defined map from $A \sq \Gc$ to $A^G$.

It should be noted that the variety $A^G$ is a simple
example of a general class of varieties that are the subject of a
deep and beautiful area of mathematics called geometric invariant
theory \cite{mumford}.  Fortunately, in the
specific case we are interested in here we can prove the desired
result without using any particularly sophisticated or delicate
methods from algebraic geometry.

\medskip
\noindent
{\bf Proof of Theorem:} We prove two basic statements, of which the
theorem is a consequence:
\begin{enumerate}
\item[(i)] For $p \in A^G$, $\pi^{-1}(p)$
contains  at most a single extended  $\Gc$ orbit.
\item[(ii)] $\pi$ is onto.
\end{enumerate}

It will be useful to define a {\em Reynolds operator} $E:R \rightarrow
R_G$, which is a projection onto the subring $R_G$ of invariants.
Because $R$ is a direct sum of finite dimensional irreducible
representations of $G$, such an operator always exists.  Important
properties of the Reynolds operator are that it is linear, and that
that for any $f \in R_G$ and $g \in R$, $E(fg) = f E(g)$.

To prove (i), we begin by noting that every extended orbit is
Z-closed.  This follows from the fact that every orbit is
constructable, which implies that the Z-closure and closure of each
orbit are identical.  Now, suppose that there were two distinct
extended orbits $O$ and $O'$ in $\pi^{-1}(p)$.  Since $O$ and $O'$ are
disjoint, the ideal $I(O) + I(O')$ in $R$ generated by $I(O)$ and
$I(O')$ must be all of $R$.  (To see this, note that the ideal $I(O) +
I(O')$ cannot be contained in any maximal ideal of $R$ or the
corresponding point would be in both $O$ and $O'$.)
%contains the largest ideals in $R$ containing $I(O)$ and
%$I(O')$; the {\em Nullstellensatz} then shows that $I(O) + I(O') = R$.)
Thus, $1 \in I(O) + I(O')$, and we can write, for some $f \in I(O)$
and $f' \in I(O')$, $1 = f + f'$.  But then we have $1=E (1) = E(f) +
E(f')$. We now claim that $E(f) \in I(O) \cap R_G$ and $E(f') \in
I(O') \cap R_G$.  This follows from the fact that the ideals $I (O)$
and $I (O')$ are invariant under $\Gc$ (since the extended orbits are
invariant) and therefore can be written as a direct sum of linear
spaces on which $\Gc$ acts irreducibly.  We have thus shown that
$E(f)$ is an invariant function that takes the value  0 on $O$ and 1 on
$O'$.  Thus, $\pi(O) \ne \pi(O')$, completing the proof of (i).

To show that $\pi$ is onto, we fix a point $p \in A^G$ and show that
there exists a nontrivial ideal $M$ in $R$ with zero set
$Z(M) = \pi^{-1}(p)$.
We define $M$ to be the ideal in $R$ generated by the maximal ideal
$I(p) \subset R_G$.  $M$ satisfies $Z(M) = \pi^{-1}(p)$ by
construction, but we must prove that $M$ is not all of $R$, so that it
is nontrivial.  To do this, note that every $g \in M$ can be written
as $g = \sum e_i f_i$, where the $\{ e_i \}$ generate $I(p)$ and $f_i
\in R$.  If $g$ is invariant, we have $g = E(g) = \sum e_i E(f_i) \in
I(p)$, which shows that $R_G \cap M = I(p)$.  Thus, $M$ is nontrivial,
proving (ii).

\section*{Acknowledgments}
We thank H. Georgi for sharing closely related work with us, and for
clarifying a useful point in our presentation.  Thanks to M. Artin and
D. Vogan for helping us to navigate the periphery of geometric
invariant theory.  We also thank M. Bershadsky, J. March--Russell, S.
Mathur, H.  Murayama, A. Nelson, L. Randall, V. Sadov, and I. Singer
for helpful conversations.  This work was supported in part by funds
provided by the U.S. Department of Energy under cooperative agreements
DE-FC02-94ER40818 and DE-AC02-76ER03069, and by the divisions of
Applied Mathematics of the U.S. Department of Energy under contracts
DE-FG02-88ER25065 and DE-FG02-88ER25066, and by National Science
Foundation grant PHY89-04035.

\end{document}